\documentclass[conference]{IEEEtran}
\usepackage{latexsym}
\usepackage{graphicx}
\usepackage{amsfonts,amssymb,amsmath}
\usepackage{algorithm,algorithmic}
\usepackage{hyperref}

\usepackage[T1]{fontenc}
\usepackage{cite}
\usepackage{subcaption}
\usepackage{comment}
\usepackage{amsthm}
\usepackage{overpic}
\usepackage{steinmetz}
\usepackage{array}
\usepackage{url}
\usepackage{color}
\IEEEoverridecommandlockouts

\theoremstyle{plain}
\newtheorem{theorem}{Theorem}

\newcommand{\argmax}[1]{{\underset{{#1}}{\mathrm{arg\,max}}}}
\newcommand{\argmin}[1]{{\underset{{#1}}{\mathrm{arg\,min}}}}
\newcommand{\vect}[1]{\mathbf{#1}}

\def\diag{\mathrm{diag}}

\def\Htran{\mbox{\tiny $\mathrm{H}$}}
\def\Ttran{\mbox{\tiny $\mathrm{T}$}}
\def\CN{\mathcal{N}_{\mathbb{C}}} 

\begin{document}

\title{Maximum Likelihood Channel Estimation for RIS-Aided Communications With LOS Channels}

\author{Emil Bj{\"o}rnson, Parisa Ramezani\\
\IEEEauthorblockA{\textit{Department of Computer Science, KTH Royal Institute of Technology, SE-100 44 Stockholm, Sweden}\\ Email: (\{emilbjo, parram\}@kth.se)}%
\thanks{This work was supported by the FFL18-0277 grant from SSF.}
}

\maketitle

\begin{abstract}
A reconfigurable intelligent surface (RIS) reflects incoming signals in different ways depending on the phase-shift pattern assigned to its elements. The most promising use case is to aid the communication between a base station and a user when the user has a line-of-sight (LOS) channel to the RIS but the direct channel is blocked. The main challenge is to estimate the channel with limited resources because non-parametric estimation methods require a pilot length proportional to the large number of RIS elements. In this paper, we develop a parametric maximum likelihood (ML) channel estimation framework for estimating the LOS channel to the RIS. We demonstrate that the proposed algorithm can accurately obtain the channel and preferred RIS configuration using only a few pilots. A key novelty is that the RIS configurations used during pilot transmission are selected to progressively improve the estimation accuracy.
\end{abstract}

\maketitle

\section{Introduction}

The propagation conditions between a base station (BS) and user equipment (UE) can be improved using a reconfigurable intelligent surface (RIS) \cite{Huang2018a}, which reflects the transmitted signal towards the receiver in a controllable manner. The potential rate and energy-efficiency benefits of the RIS technology are well understood but a remaining practical challenge is channel estimation \cite{Bjornson2020a}.
While the estimation in conventional systems is limited by the signal-to-noise ratio (SNR), the largest issue in RIS-aided communications is huge pilot dimensionality.
Both the least-squares estimator \cite{Jensen2020a} and minimum mean-squared error estimator \cite{Wang2020a} require a pilot length proportional to the number of RIS elements, and hundreds of elements are needed for RIS to outperform relays \cite{Bjornson2019e}. The array geometry can be utilized to make the pilot length proportional to the RIS area \cite{Demir2022a}, which is smaller for dense RIS arrays but still large.

The frequency response of an RIS element is flat, thus it can only control the propagation over a (nearly) narrowband channel \cite{Bjornson2022a}. Current and future systems use wide bandwidths, but are sufficiently narrowband if there are line-of-sight (LOS) channels to and from the RIS. In contrast, \cite[Fig.~6]{Bjornson2022a} shows that an RIS provides no appreciable gains over wideband non-LOS channels.
This insight can be exploited to simplify the channel estimation in the practically important LOS scenario.
A review of parametric channel estimation for RIS-aided systems is provided in \cite{Swindlehurst2022a}.
A parametric maximum likelihood (ML) estimator is developed in \cite{Wang2021a} for the case where both the LOS channels to and from the RIS are unknown. A beam-training procedure for LOS channels is developed in \cite{Ning2021a}.
A common property of these estimation methods is that a full-dimensional search over the RIS configurations is made.

In this paper, we consider the estimation of the LOS channel between the UE and RIS, while the channel between the RIS and BS is assumed known from the network deployment.
We derive a parametric ML estimator following the classical deterministic ML approach \cite{Krim1996a}.
As a key novelty, we select the RIS configurations during pilot transmission to progressively refine the estimation accuracy.
This enables us to use much fewer pilots than when all dimensions are explored, as in previous non-parametric \cite{Jensen2020a,Wang2020a} and parametric estimation \cite{Wang2021a,Ning2021a} for RIS systems. The effectiveness is demonstrated numerically.

\section{System Model}
\label{sec:sysmod}

We consider an RIS-assisted communication system, where the transmission between the single-antenna UEs and
single-antenna BS is aided by an RIS consisting of $N$ reconfigurable elements.
The RIS and BS are deployed at fixed locations, thus the channel vector $\vect{h} = [h_1,\ldots,h_N]^{\Ttran} \in \mathbb{C}^N$ between them is assumed to be known. 
The BS serves a multitude of UEs at different locations in time-division multiple access (TDMA) mode. Hence, the UE channel must be estimated every time a UE is scheduled and the RIS must be configured based on the new estimate.
We let the channel between an arbitrary UE and the RIS be denoted by $\vect{g} = [g_1,\ldots,g_N]^{\Ttran}  \in \mathbb{C}^N$.
To focus on the estimation of this channel, we assume there is no direct link between the BS and UE.

If the UE transmits the signal $x \in \mathbb{C}$ to the BS via the RIS, the received signal can be expressed as \cite{Bjornson2022a}
\begin{equation}
\label{eq:received_signal}
 y = \left( \sum_{n=1}^{N} h_n g_n e^{-j\theta_n} \right) x + w, 
\end{equation}
$\theta_n$ is the phase-shift induced by the $n$th RIS element for $n=1,\ldots,N$. 
If a data signal $x \sim \CN(0,P_d)$ is transmitted with power $P_d$, the capacity of this channel is \cite[Lemma~1]{Bjornson2019e}
\begin{align} \label{eq:rate-expression}
    &\log_2 \!\left( 1 + \left| \sum_{n=1}^{N} h_n g_n e^{-j\theta_n} \right|^2 \frac{P_d }{\sigma^2} \right) \\ &\leq \log_2 \!\left( 1 + \left( \sum_{n=1}^{N} |h_n g_n| \right)^2 \!\frac{P_d}{\sigma^2} \right), \label{eq:capacity-expression}
\end{align}
where the former expression holds for any RIS phase-shift configuration and the upper bound is achieved by selecting the phase-shifts as $\theta_n = \arg(h_n)+\arg(g_n)$ for $n=1,\ldots,N$. 
This optimal selection makes $ h_n g_n e^{-j\theta_n} = |h_n g_n| $ so that all the $N$ propagation paths via the RIS are phase-aligned.

To use this optimal configuration, the system needs to know $\arg(h_n)+\arg(g_n)$ for all elements so channel state information (CSI) is essential. Since $\vect{h}$ is assumed known, the problem is to estimate the channel $\vect{g}$ between the UE and RIS. 

\section{Maximum Likelihood Channel Estimation}

When considering estimation of the channel $\vect{g}$, it is convenient to rewrite \eqref{eq:received_signal} in the equivalent form
\begin{equation}
   y = \boldsymbol{\theta}^{\Ttran} \vect{D}_{\vect{h}} \vect{g} x + w,
\end{equation}
by defining the vector $\boldsymbol{\theta} = \left[e^{-j\theta_1},\ldots,e^{-j\theta_N}\right]^{\Ttran} \in \mathbb{C}^N$ containing the RIS phase-shifts and the diagonal matrix  $\vect{D}_{\vect{h}}= \diag (h_1,\ldots,h_N)$ with the channel coefficients between the RIS and BS.
The effective scalar channel $\boldsymbol{\theta}^{\Ttran} \vect{D}_{\vect{h}} \vect{g}$ is a projection of $\vect{g}$ onto the direction of the vector $\boldsymbol{\theta}^{\Ttran} \vect{D}_{\vect{h}}$ determined by the RIS configuration.

To estimate the unknown channel $\vect{g}$ at the BS, the UE needs to transmit a known pilot signal and the RIS needs to switch between different configurations so that the projections of $\vect{g}$ in different directions are observable.
Suppose the UE transmits the deterministic pilot signal $x = \sqrt{P_p}$, with power $P_p$, at $L$ time instances.
The RIS configuration is changed between these instances as $\boldsymbol{\theta}_1,\ldots,\boldsymbol{\theta}_L$.
The concatenated received signal $\vect{y} \in \mathbb{C}^N$ at the BS can then be expressed as  
\begin{equation} \label{eq:received-pilot}
    \vect{y} = \vect{B} \vect{D}_{\vect{h}} \vect{g} \sqrt{P_p} + \vect{w},
\end{equation} 
where
\begin{align}
   \label{eq:configuration_matrix} \vect{B} &= \left[\boldsymbol{\theta}_1,\ldots,\boldsymbol{\theta}_L\right]^{\Ttran}, \\
    \vect{w} &= [w_1,\ldots,w_L]^{\Ttran},
\end{align} 
and $w_l$ is the noise at pilot time instance $l$, for  $l=1,\ldots,L$.

There is a multitude of channel estimators that can be developed based on the received pilot signal in \eqref{eq:received-pilot}.
We will take the ML estimation approach \cite{Kay1993a} because the channel $\vect{g}$ is assumed unknown but deterministic. 
The probability density function (PDF) of $\vect{y}$ for a given $\vect{g}$ can be expressed as 
\begin{equation}
\label{eq:y_PDF}
    f_{\vect{Y}}(\vect{y}; \vect{g}) = \frac{1}{(\pi \sigma ^2)^L} e^{- \frac{\|\vect{y} - \vect{B}\vect{D}_{\vect{h}} \vect{g} \sqrt{P_p}\|^2}{\sigma^2}}.
\end{equation} 
Due to the ML criterion, we are looking for the channel estimate $\hat{\vect{g}}$ that maximizes the PDF in \eqref{eq:y_PDF}:
\begin{align}
\label{eq:g_hat_general}
    \hat{\vect{g}} = \argmax{\vect{g} \in \mathbb{C}^N}~f_{\vect{Y}}(\vect{y}; \vect{g}) =  \argmin{\vect{g} \in \mathbb{C}^N}~ \left\|\vect{y} - \vect{B}\vect{D}_{\vect{h}} \vect{g} \sqrt{P_p} \right\|^2.
\end{align}
If all vectors in $\mathbb{C}^N$ are plausible channels, then the solution to \eqref{eq:g_hat_general} is
\begin{equation} \label{eq:ML-general-L}
    \hat{\vect{g}} = \frac{1}{\sqrt{P_p}} \vect{D}_{\vect{h}}^{-1} \vect{B}^{\dagger} \vect{y}
\end{equation}
where $\vect{B}^{\dagger}$ denotes the pseudoinverse of $\vect{B}$.
If $L \geq N$ and $\vect{B}$ is selected to have rank $N$, then \eqref{eq:ML-general-L} simplifies to
\begin{equation} \label{eq:ML-general-L-simplified}
    \hat{\vect{g}} = \frac{1}{\sqrt{P_p}} \vect{D}_{\vect{h}}^{-1} (\vect{B}^{\Htran} \vect{B})^{-1} \vect{B}^{\Htran} \vect{y}.
\end{equation}
Apart from the fact that $\vect{D}_{\vect{h}}$ is assumed known, \eqref{eq:ML-general-L-simplified} is basically the same ML estimator as in \cite{Swindlehurst2022a}, which can also be called the minimum variance unbiased estimator, or least-squares estimator \cite{Bjornson2022a}. We will use the latter terminology herein.
The main practical issue is that we need $L \geq N$ to utilize \eqref{eq:ML-general-L-simplified}, which requires very long pilot sequences when considering a typical RIS with hundreds of elements \cite{Bjornson2019e}.
The estimator in \eqref{eq:g_hat_general} can be used for the desired case of $L < N$ but it will not perform well since there are fewer observations than unknowns.
However, we will now demonstrate that the ML framework becomes very powerful when combined with a known channel parameterization that reduces the set of feasible channel realizations to a small subset of $\mathbb{C}^N$.

\vspace{-1mm}
\subsection{Parametric ML Estimation}

Suppose only channel vectors $\vect{g} \in \mathcal{A}$ are feasible in the considered scenario, where $\mathcal{A} \subset \mathbb{C}^N$ is a subset of all possible vectors.
In the considered LOS scenario, we can assume
\begin{equation}
    \mathcal{A} = \{ c \, \vect{a} ({\varphi}): c \in \mathbb{C}, \varphi \in \Phi \},
\end{equation}
where $c$ can be any complex channel coefficient at the reference element of the RIS and $\vect{a} ({\varphi})$ is the array response vector for a plane wave arriving from the angle-of-arrival (AOA) $\varphi$. The array response vector is a function of the AOA but is determined by the array geometry so the function is known. The set $\Phi$ of feasible AOAs is also deployment-specific.

We can utilize this parametrization to particularize \eqref{eq:g_hat_general} as the new \emph{parametric ML estimator}
\begin{align}
\label{eq:g_hat_parametric}
    \hat{\vect{g}} =  \hat{c} \, \vect{a} (\hat{\varphi}) =  \argmin{c \in \mathbb{C}, \varphi \in \Phi}~ \left\|\vect{y} - c \vect{B}\vect{D}_{\vect{h}} \vect{a} ({\varphi}) \sqrt{P_p} \right\|^2.
\end{align}
We further introduce the notation $c=\sqrt{\beta}e^{j\omega}$, where $\beta \geq 0 $ is the channel gain and $\omega$ is the phase-shift at the reference element. By this change of variables, \eqref{eq:g_hat_parametric} can be rewritten as a joint estimation of $\beta$, $\omega$, and $\varphi$: \vspace{-2mm}
\begin{align}
\label{eq:channel_estimation_1}
    &\{\hat{\beta},\hat{\omega},\hat{\varphi}\} = \argmin{\substack{\beta \geq 0,\omega \in [0,2\pi),\\ \varphi \in \Phi}} ~\left|\left|\vect{y} - \sqrt{\beta} e^{j\omega}\vect{B}\vect{D}_{\vect{h}} \vect{a}(\varphi) \sqrt{P_p}\right|\right|^2 \notag\\ &= 
    \argmin{\substack{\beta \geq 0,\omega \in [0,2\pi),\\ \varphi \in \Phi}} ~P_p\beta \left|\left|\vect{B}\vect{D}_{\vect{h}}\vect{a}(\varphi)\right|\right|^2 - 2 \sqrt{P_p\beta}\text{Re}\left(e^{j\omega} \vect{y}^{\Htran}\vect{B} \vect{D}_{\vect{h}} \vect{a}(\varphi)\right),
\end{align} \vspace{-3mm}

\noindent where the term $\|\vect{y}\|^2$ is omitted in the second line since it is independent of the optimization variables. 

First, we can notice that $\omega$ only appears in the last term of \eqref{eq:channel_estimation_1}. We, therefore, need to maximize $\text{Re}\left(e^{j\omega} \vect{y}^{\Htran}\vect{B} \vect{D}_{\vect{h}} \vect{a}(\varphi)\right)$ to find the ML estimator, from which $\hat{\omega}$ is obtained as 
\begin{equation}
\label{eq:theta_estimate}
    \hat{\omega} = -\mathrm{arg}\left(\vect{y}^{\Htran}\vect{B} \vect{D}_{\vect{h}} \vect{a}(\varphi)\right),
\end{equation}
which makes  
$\text{Re}\left(e^{j\omega} \vect{y}^{\Htran}\vect{B} \vect{D}_{\vect{h}} \vect{a}(\varphi)\right) = |\vect{y}^{\Htran}\vect{B} \vect{D}_{\vect{h}} \vect{a}(\varphi)|$.
Substituting this expression into \eqref{eq:channel_estimation_1} yields 
\begin{equation}
\label{eq:channel_estimation_2}
    \{\hat{\beta},\hat{\varphi}\} = \argmin{\beta \geq 0,\varphi \in \Phi} ~P_p\beta \left|\left|\vect{B}\vect{D}_{\vect{h}}\vect{a}(\varphi)\right|\right|^2 - 2 \sqrt{P_p\beta}\left|\vect{y}^{\Htran}\vect{B} \vect{D}_{\vect{h}} \vect{a}(\varphi)\right|.
\end{equation}
This is a second-order polynomial with respect to 
$\sqrt{\beta}$. By identifying the only non-zero root,
 $\hat{\beta}$ is readily obtained as 
\begin{equation}
\label{eq:beta_estimate}
    \hat{\beta} =  \frac{\left|\vect{y}^{\Htran}\vect{B} \vect{D}_{\vect{h}} \vect{a}(\varphi)\right|^2}{P_p \left|\left|\vect{B}\vect{D}_{\vect{h}}\vect{a}(\varphi)\right|\right|^4}.
\end{equation}
Substituting \eqref{eq:beta_estimate} into \eqref{eq:channel_estimation_2}, the ML estimate for the AOA can be found as 
\begin{align}
   \hat{\varphi} &= \argmin{\varphi \in \Phi}~- \frac{\left|\vect{y}^{\Htran}\vect{B} \vect{D}_{\vect{h}} \vect{a}(\varphi)\right|^2}{\left|\left|\vect{B}\vect{D}_{\vect{h}}\vect{a}(\varphi)\right|\right|^2} \notag \\
   & =\argmax{\varphi \in \Phi}~ \frac{\left|\vect{y}^{\Htran}\vect{B} \vect{D}_{\vect{h}} \vect{a}(\varphi)\right|^2}{\left|\left|\vect{B}\vect{D}_{\vect{h}}\vect{a}(\varphi)\right|\right|^2} . \label{eq:ML-estimate}
\end{align}
This maximization needs to be solved numerically. The utility function is continuous but it is known in the array signal processing literature that ML problems of this kind can have many local maxima \cite{Krim1996a}.
There are methods to precisely identify all of them \cite{Wang2021a}, which is necessary if their peak values are similar. However, in our application, a grid search is sufficient since Section~\ref{subsec:RIS-configuration-estimation} develops an algorithm to gradually improve the utility function until there is only one distinct peak value.
Note that the same received signal $\vect{y}$ is utilized every time the utility is evaluated, thus the granularity of the grid search only affects the computational complexity.

In summary, we have proved the following main result.

\begin{theorem} \label{th:main-result}
The parametric ML estimate $\hat{\vect{g}} = \hat{c} \, \vect{a} (\hat{\varphi})$ is
\begin{align}
    \hat{c} &= \sqrt{\hat{\beta}} e^{j \hat{\omega}}= 
 \frac{\left|\vect{y}^{\Htran}\vect{B} \vect{D}_{\vect{h}} \vect{a}(\hat{\varphi})\right|}{\sqrt{P_p} \left|\left|\vect{B}\vect{D}_{\vect{h}}\vect{a}(\hat{\varphi})\right|\right|^2}
    e^{-j \mathrm{arg}\left(\vect{y}^{\Htran}\vect{B} \vect{D}_{\vect{h}} \vect{a}(\hat{\varphi})\right)}, \\
   \hat{\varphi} & =\argmax{\varphi \in \Phi}~ \frac{\left|\vect{y}^{\Htran}\vect{B} \vect{D}_{\vect{h}} \vect{a}(\varphi)\right|^2}{\left|\left|\vect{B}\vect{D}_{\vect{h}}\vect{a}(\varphi)\right|\right|^2} .
\end{align}
\end{theorem}

This ML estimator can be interpreted as follows. The AOA is estimated by computing  the inner product between the actually received signal $\vect{y}$ and the normalized noise-free received signal $\vect{B} \vect{D}_{\vect{h}} \vect{a}(\varphi) / \| \vect{B} \vect{D}_{\vect{h}} \vect{a}(\varphi) \|$, and 
identifying which feasible angle $\varphi \in \Phi$ maximizes its squared magnitude.

The estimated received signal $\vect{B} \vect{D}_{\vect{h}} \vect{a}(\hat{\varphi}) $ lacks the scaling factor $c=\sqrt{\beta} e^{j \omega}$.
The phase-shift $\omega$ is estimated by compensating for the phase difference between the received signal and channel direction
$\vect{B} \vect{D}_{\vect{h}} \vect{a}(\hat{\varphi}) $.
Finally, the channel magnitude $\sqrt{\beta}$ is estimated by computing the same inner product and then normalizing it so that only the magnitude remains.

\subsection{Example: Uniform linear array}

Suppose the RIS is a horizontally deployed uniform linear array (ULA) with $N$ elements and spacing $\Delta$. 
The array response vector at the wavelength $\lambda$ can be expressed as
\begin{equation}
\label{eq:channel_ULA}
    \vect{a}_{\textrm{ULA}} ({\varphi}) = 
\begin{bmatrix} 1 \\ e^{-j2\pi \frac{\Delta \sin(\varphi)}{\lambda}} \\ \vdots  \\  e^{-j2\pi \frac{(N-1)\Delta \sin(\varphi)}{\lambda}} \end{bmatrix},
\end{equation}
where $\varphi$ is the azimuth AOA. The RIS can only receive signals in front of it, thus the ML estimate in \eqref{eq:ML-estimate} becomes
\begin{align}
   \hat{\varphi}  =\argmax{\varphi \in [-\frac{\pi}{2},\frac{\pi}{2}]}~ \frac{\left|\vect{y}^{\Htran}\vect{B} \vect{D}_{\vect{h}} \vect{a}_{\textrm{ULA}} (\varphi)\right|^2}{\left|\left|\vect{B}\vect{D}_{\vect{h}}\vect{a}_{\textrm{ULA}} (\varphi)\right|\right|^2}. \label{eq:ML-ULA}
\end{align}
The same expression can be used if the RIS is planar but deployed at the same height as the prospective UEs so that only the azimuth angle determines the channel. If the RIS has $M$ rows and $N$ elements per row, then $\vect{a}_{\textrm{ULA}}$ is the array response of each row. By assigning the same phase shift to all elements in a column, we can reduce the planar RIS problem to the one studied in this paper but with $M^2$ times larger SNR.

\section{Adaptive RIS Configuration During Estimation}
\label{subsec:RIS-configuration-estimation}

The RIS configuration matrix $\vect{B} = \left[\boldsymbol{\theta}_1,\ldots,\boldsymbol{\theta}_L\right]^{\Ttran}  \in \mathbb{C}^{L \times N}$ determines the quality of the parametric ML estimation procedure described above. We will now consider how to select this matrix.
We would like the number of RIS configurations to be much fewer than the number of RIS elements (i.e., $L\ll N$) by only exploring important channel dimensions in the pilot transmission.
The previous approaches in 
\cite{Jensen2020a,Wang2020a,Wang2021a,Ning2021a} require $L \geq N$
since $\vect{B}$ is selected a priori or randomly.
Herein, we propose an iterative algorithm where the RIS configurations during the pilot transmission are selected adaptively to gradually refine the ML estimate. We show in Section~\ref{sec:results} that it quickly converges to a good AOA estimate.

Suppose the exact AOA from the UE to the RIS is $\bar{\varphi}$. The optimal RIS phase-shift vector, denoted by $\bar{\boldsymbol{\theta}}$, for achieving the capacity in \eqref{eq:capacity-expression} is given by
\begin{equation}
\label{eq:optimal_config}
    \bar{\boldsymbol{\theta}} = \diag \left(e^{-j\mathrm{arg}(h_1)},\ldots,e^{-j\mathrm{arg}(h_N)} \right) \vect{a}^*(\bar{\varphi}),
\end{equation}
where $^*$ denotes complex conjugation.
Hence, when $\bar{\varphi}$ is to be estimated, it is natural to consider $L$ RIS configurations with the structure in \eqref{eq:optimal_config} but using other angles.
We will select a set $\mathcal{B}= \{ \bar{\varphi}_1,\bar{\varphi}_2,\ldots,\bar{\varphi}_N \}$ of $N$ well-separated plausible angles and define the resulting set of plausible RIS configurations as
\begin{equation}
\label{eq:configuration_set}
 \Theta = \left\{ \diag \left(e^{-j\mathrm{arg}(h_1)},\ldots,e^{-j\mathrm{arg}(h_N)} \right) \vect{a}^*(\varphi) : \varphi \in \mathcal{B} \right \}.  
\end{equation}

Algorithm~\ref{Alg:AoA_Estimation} describes our proposed iterative approach for estimating the AOA between the UE and the RIS. We start by selecting two RIS configurations from the set $\Theta$ in \eqref{eq:configuration_set} to create the initial matrix $\vect{B}_2 \in \mathbb{C}^{2 \times N}$. 
Pilots are transmitted at two time instances and the received signal $\vect{y}_2 \in \mathbb{C}^{2}$ is used to compute a first ML estimate $\hat{\varphi}_2$ using Theorem~\ref{th:main-result}.
Then, an adaptive procedure begins where the next RIS configuration is selected as the unused configuration in $\Theta$ that is closest to what the optimal RIS phase-shift vector in \eqref{eq:optimal_config} would become if the ML estimate was exact. A new pilot is then transmitted using this configuration and a new ML estimate is computed using all the received signals.
This procedure is continued until the intended pilot length $L$ is reached.

The intuition behind this algorithm is that each pilot that is transmitted (after the initial two) is used to refine the current ML estimate. By using an RIS configuration that resembles the current estimate but has not yet been considered, we will either improve the certainty of the estimate or discard it in light of the new pilot signal and find a very different estimate. The algorithm can be initialized by selecting two configurations in $\Theta$ that are far apart. 
The algorithm terminates when $L$ pilots have been used for channel estimation, however, it can possibly be terminated early if the difference between the largest and second-largest peaks of the utility function is big.

\begin{algorithm}
\caption{AOA estimation between the UE and RIS.}
\label{Alg:AoA_Estimation}
\begin{algorithmic}[1]
\STATE{Define the set $\Theta$ of plausible RIS configurations in \eqref{eq:configuration_set}}
\STATE{Select two initial RIS configurations $\boldsymbol{\theta}_1$ and $\boldsymbol{\theta}_2$ from $\Theta$}
\STATE{Set $\vect{B}_2 = [\boldsymbol{\theta}_1,\boldsymbol{\theta}_2]^{\Ttran}$ and update $\Theta \gets \Theta \setminus \{\boldsymbol{\theta}_1,\boldsymbol{\theta}_2\}$}
\STATE{Send pilot signals using the RIS configurations $\boldsymbol{\theta}_1,\boldsymbol{\theta}_2$ to get the received signal $\vect{y}_2 \in \mathbb{C}^2$}
 \FOR{$i = 2,\ldots,L$}
    \STATE{Compute ML estimate $\hat{\varphi}_i  =\argmax{\varphi \in \Phi}~ \frac{\left|\vect{y}_i^{\Htran}\vect{B}_i \vect{D}_{\vect{h}} \vect{a}(\varphi)\right|^2}{\left|\left|\vect{B}_i\vect{D}_{\vect{h}}\vect{a}(\varphi)\right|\right|^2}$}
    \IF{$i < L$}
    \STATE{ Set $\bar{\boldsymbol{\theta}}_i = \diag \left(e^{-j\mathrm{arg}(h_1)},\ldots,e^{-j\mathrm{arg}(h_N)} \right) \vect{a}^*(\hat{\varphi}_i)$}
    \STATE{Compute $\boldsymbol{\theta}_{i+1} = \argmax{\boldsymbol{\theta} \in \Theta}~ | \bar{\boldsymbol{\theta}}_i^{\Htran} \boldsymbol{\theta}|$}
    \STATE{Set $\vect{B}_{i+1} = [\vect{B}_i^{\Ttran}, \boldsymbol{\theta}_{i+1}]^{\Ttran}$, update $\Theta \gets \Theta \setminus \{\boldsymbol{\theta}_{i+1}\}$}
    \STATE{Send a pilot signal using the RIS configuration $\boldsymbol{\theta}_{i+1}$ and collect received signals in $\vect{y}_{i+1} = [\vect{y}_{i}^{\Ttran},y_{i+1}]^{\Ttran}$}
\ENDIF
 \ENDFOR 
\RETURN ML estimate $\hat{\varphi}_L$
 \end{algorithmic}
\end{algorithm}

\subsection{Example: Uniform linear array}

If a ULA with the array response vector in \eqref{eq:channel_ULA} is used, then the magnitude of the inner product between the vectors for two different angles $\bar{\varphi}_{n_1},\bar{\varphi}_{n_2} \in \mathcal{B}$ becomes
\begin{align}
\label{eq:beam_angles}
|\vect{a}_{\mathrm{ULA}}^{\Htran}(\bar{\varphi}_{n_1})\vect{a}_{\mathrm{ULA}}(\bar{\varphi}_{n_2})| =  \frac{\sin \left(\frac{N\pi \Delta}{\lambda} \left(\sin(\bar{\varphi}_{n_2}) - \sin(\bar{\varphi}_{n_1})\right) \right)}{\sin \left(\frac{\pi \Delta}{\lambda} \left(\sin(\bar{\varphi}_{n_2}) - \sin(\bar{\varphi}_{n_1})\right) \right)}.
\end{align}
The value is determined by the difference between $\sin(\bar{\varphi}_{n_1})$ and $\sin(\bar{\varphi}_{n_2})$, thus the plausible angles in $\mathcal{B}$ are well separated if the sine of them are equally spaced between $-1$ and $+1$.
Hence, we select the set of plausible angles as
\begin{equation}
\label{eq:angles_set}
 \mathcal{B} = \left\{\mathrm{arcsin} \left(\frac{2m}{N}\right) : m = - \left \lfloor \frac{N-1}{2} \right \rfloor, \ldots, 0, \ldots, \left \lfloor \frac{N}{2} \right \rfloor \right \},   
\end{equation} 
where $\lfloor \cdot \rfloor$ truncates the argument to the closest smaller integer.
The plausible RIS configurations can now  be computed by \eqref{eq:configuration_set} using the array response vector for ULAs and $\mathcal{B} $ in \eqref{eq:angles_set}.

\section{Numerical Results}
\label{sec:results}

We will now demonstrate the effectiveness of the proposed parametric ML estimator and adaptive RIS configuration. We consider a setup where the RIS and UE are located in the same horizontal plane, so that only the azimuth AOA must be estimated. The RIS is rectangular and has $N=40$ horizontal elements that are separated by $\Delta=\lambda/4$. The number of vertical elements will not affect our results so we leave it arbitrary.
There is a known LOS channel between the BS and RIS, and the choice of angular direction will not affect the results since we compensate fully for $\vect{D}_{\vect{h}} $ in the RIS configuration.
The UE locations are selected uniformly at random from $[-\pi/3,\pi/3]$, while $\Phi = [-\pi/2,\pi/2]$.

We define the per-element data SNR as 
\begin{equation}
   \mathrm{SNR}_d = \frac{P_d}{\sigma^2} | h_n g_n |^2,
\end{equation}
which is independent of the element index $n$ since we consider LOS channels. 
We further assume that the pilot SNR, $\mathrm{SNR}_p$, is $10$\,dB larger than $\mathrm{SNR}_d$ (i.e., $P_p = 10 P_d$), motivated by the fact that a pilot processing gain can be achieved by spreading the pilot sequence of the frequency domain.

\begin{figure}[t!]
        \centering
        \begin{subfigure}[b]{\columnwidth} \centering
	\begin{overpic}[width=1.05\columnwidth,tics=10]{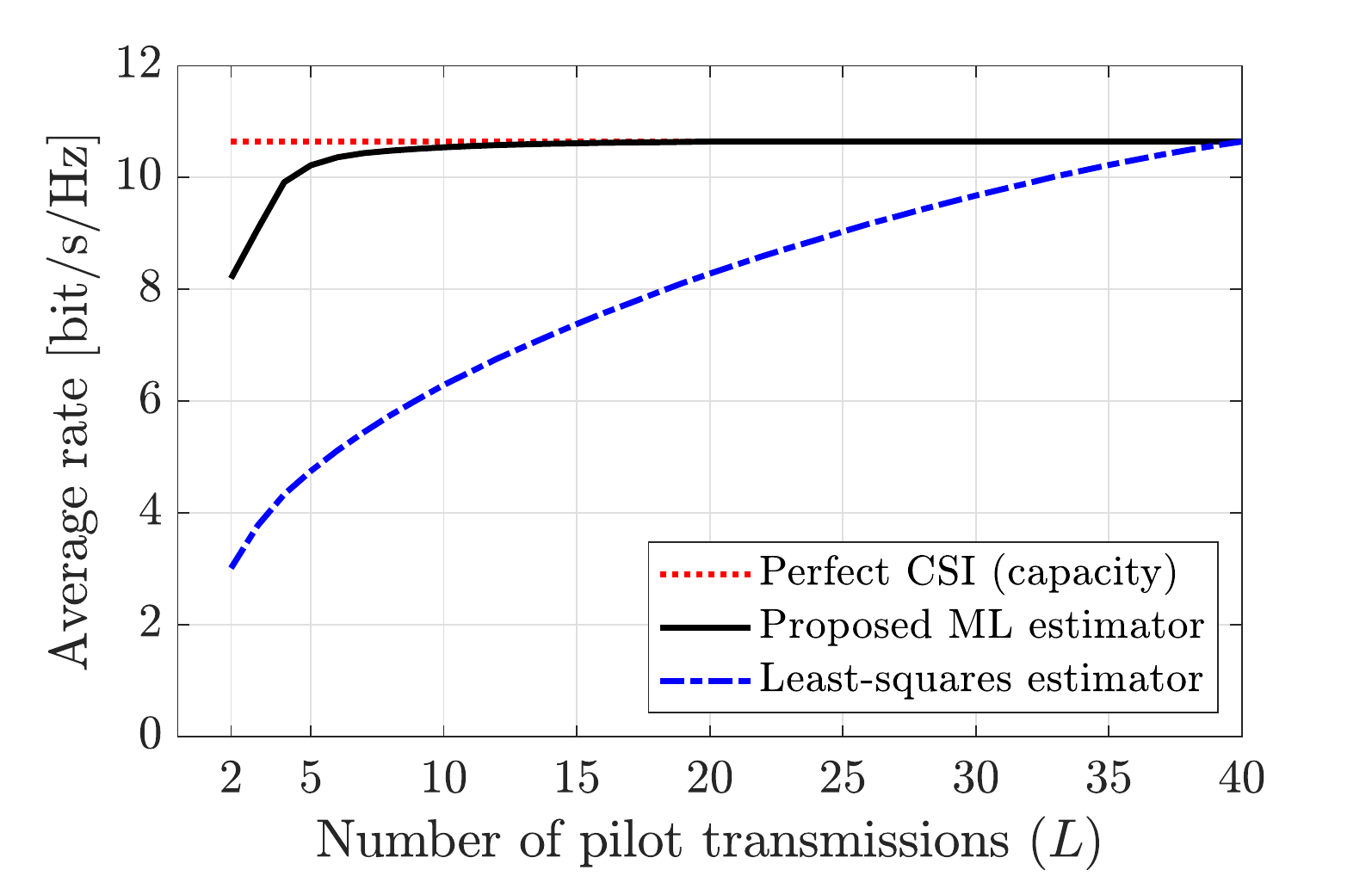}
\end{overpic} 
                \caption{$\mathrm{SNR}_p=10$\,dB, $\mathrm{SNR}_d=0$\,dB.}
    \vspace{+2mm}
        \end{subfigure} 
        \begin{subfigure}[b]{\columnwidth} \centering
	\begin{overpic}[width=1.05\columnwidth,tics=10]{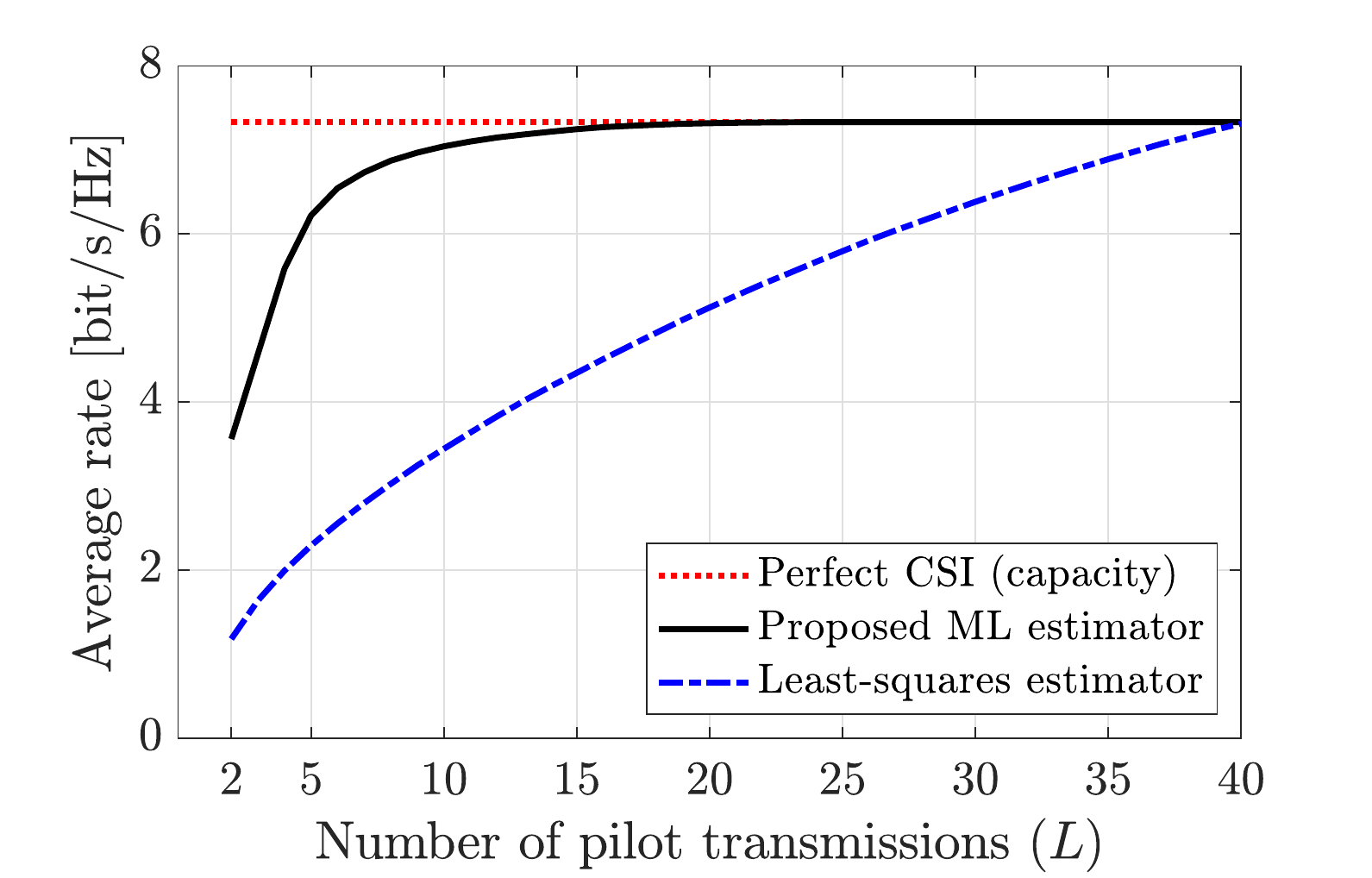}
\end{overpic} 
                \caption{$\mathrm{SNR}_p=0$\,dB, $\mathrm{SNR}_d=-10$\,dB.} 
        \end{subfigure} 
        \caption{The average rate (over different AOA and noise realizations) as a function of the number of pilot transmissions $L$ with different RIS configurations. The proposed parametric ML estimator is compared with the perfect CSI case and least-squares estimator considered in prior work.} 
        \label{fig:simulationRates}  
\end{figure}

Fig.~\ref{fig:simulationRates} shows the average rate (over different AOA and noise realizations) as a function of the number of pilot transmissions, $L$, with different RIS configurations during the channel estimation (implemented according to Section~\ref{subsec:RIS-configuration-estimation}).
The rate achieved when configuring the RIS based on the proposed parametric ML estimate of the AOA is computed using \eqref{eq:rate-expression}. We compare it with the capacity expression in \eqref{eq:capacity-expression} that is achieved with perfect CSI. We also benchmark the proposed algorithm against the conventional least-squares estimator in \eqref{eq:ML-general-L}, in which case a random subset of the columns of a DFT matrix is used as the
$L$ RIS configurations during the pilot transmission.

Fig.~\ref{fig:simulationRates}(a) considers $\mathrm{SNR}_p = 10$\,dB. The proposed algorithm achieves an average of $76$\% of the capacity for $L=2$ pilot transmissions (and RIS configurations). The percentage then increases rapidly with $L$ as the ML estimate is gradually refined by sending extra pilots using RIS configurations that seemed promising. $93$\% of the capacity is achieved at $L=4$ and $96\%$ at $L=5$. It is clear that the proposed algorithm finds a good estimate of $\vect{g}$ using much less than $N=40$ pilot transmissions. In contrast, the conventional least-squares estimator begins at a much lower average rate and requires $L \approx N$ to come close to the capacity.

Fig.~\ref{fig:simulationRates}(b) considers the case of $\mathrm{SNR}_p = 0$\,dB. The reduced SNR results in a lower capacity, but also worse estimation quality. The proposed ML estimator is more affected by the noise since it relies on comparing the received signal with prospective noiseless signals to identify the most likely one. Nevertheless, the proposed algorithm greatly outperforms the least-squares estimation benchmark and $L=10$ is sufficient to reach $96\%$ of the capacity, instead of $L \approx 40$.

\begin{figure}[t!]
        \centering
	\begin{overpic}[width=\columnwidth,tics=10]{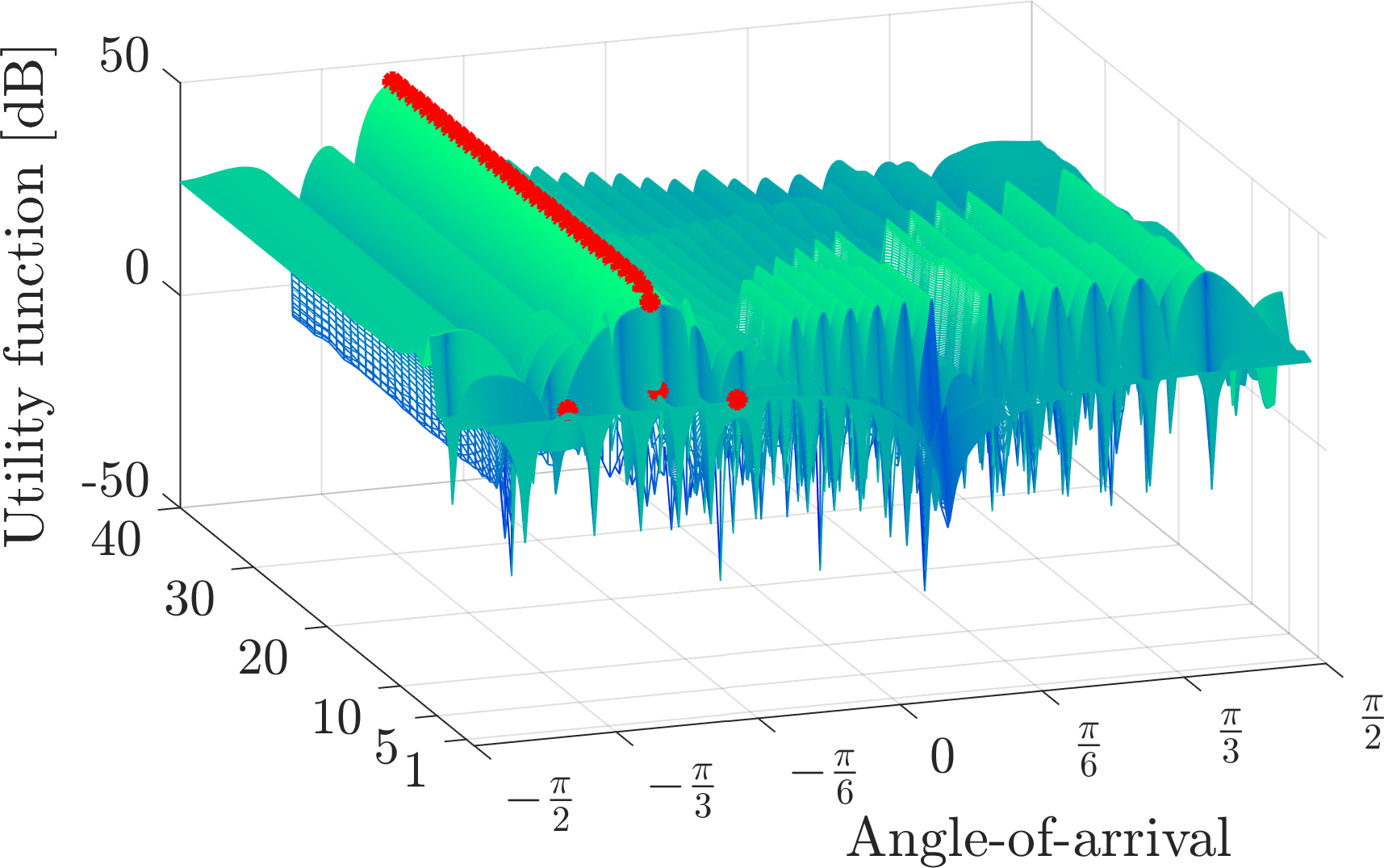}
	\put(2,6.2){Number of}
	\put(2,2){pilots ($L$)}
\end{overpic} 
        \caption{Illustration of the utility function in \eqref{eq:ML-estimate} and how it evolves with the number of pilot transmissions $L$. The peak values (marked with red stars) are the ML estimates.} 
        \label{fig:simulationIllustration}  
\end{figure}

The typical convergence behavior is illustrated in Fig.~\ref{fig:simulationIllustration}. The utility function in \eqref{eq:ML-estimate} is shown (in the decibel scale due to the large dynamic range) for different AOAs and number of pilot transmissions $L$. The red stars illustrate the ML estimate for each value of $L$ and the true value is $\varphi=-\pi/4$.
In the first few iterations ($L<5$), the utility function has many local maxima with similar peak values. Hence, multiple AOAs are roughly equally likely and there is a large risk that the wrong one is selected for the ML estimate. 
Every time $L$ is increased, a pilot is transmitted using an RIS configuration that resembles the last ML estimate, according to the algorithm proposed in Section~\ref{subsec:RIS-configuration-estimation}. If the previous estimate was far from the true AOA, then the new pilot transmission will help to exclude this angle; all the peaks of the utility function diminish except the one at the true AOA. As can be seen in the figure, the red stars jump between three different points before stabilizing around the true angle when $L=5$.
Depending on how similar the RIS configurations at the two initial pilot transmissions happen to be to the optimum, the initial process of excluding peaks at incorrect AOAs can take $2$-$10$ iterations. 
Hence, if the system has any prior information about where the UE might be located, the algorithm could be initialized based on this information to speed up convergence.

\section{Conclusions}

Accurate CSI is essential for an RIS to improve the channel quality between a transmitter and receiver. Recent works have shown that LOS channels to and from the RIS are needed to support wide bandwidths. In this paper, we have developed an efficient channel estimation algorithm that exploits the channel structure in this scenario. It consists of a parametric ML estimator and an adaptive way of selecting which RIS configurations to use during pilot transmission to gradually refine the estimate. This novel algorithm achieves accurate estimates using much fewer pilots than there are RIS elements.

\bibliographystyle{IEEEtran}
\bibliography{IEEEabrv,refs}

\end{document}